\documentclass[11pt,a4paper]{article}
\pdfoutput=1
\usepackage{jheppub}
\usepackage{amssymb,amsmath,amstext,amsfonts,empheq}
\usepackage{bbold}
\counterwithout{equation}{section} 
\usepackage{simpler-wick}
\usepackage{bm}
\usepackage{graphicx}
\usepackage{xcolor}
\usepackage{overpic}
\usepackage{subcaption}
\usepackage{braket}
\usepackage[extra]{tipa}

\newcommand{\be}{\begin{equation}}
\newcommand{\ee}{\end{equation}}
\newcommand{\bea}{\begin{eqnarray}}
\newcommand{\eea}{\end{eqnarray}}

\newcommand{\bal}{\begin{aligned}}
\newcommand{\eal}{\end{aligned}}

\newcommand{\Mp}{M_{\rm Pl}}

\newcommand{\R}{\zeta}

\newcommand{\Rh}{\hat{\zeta}}

\newcommand{\q}{|\boldsymbol{p}-\boldsymbol{k}|}

\newcommand{\bk}{\boldsymbol{k}}
\newcommand{\bK}{\boldsymbol{K}}
\newcommand{\bp}{\boldsymbol{p}}
\newcommand{\bq}{\boldsymbol{q}}
\definecolor{alizarin}{rgb}{0.82, 0.1, 0.26}

\newcommand{\p}{\mathcal{P}}

\usepackage{soul}
\definecolor{lgray}{gray}{0.90}

\newcommand{\bx}{\boldsymbol{x}}
\definecolor{alizarin}{rgb}{0.82, 0.1, 0.26}

\newcommand{\taue}{\tau_{\rm e}}
\newcommand{\taus}{\tau_{\rm s}}

\title{  \centering
\boldmath
Absence of one-loop effects on large scales from small scales in non-slow-roll dynamics }

\author{\Large Jacopo Fumagalli\\} 
\affiliation{Departament de F\'isica Quàntica i Astrofísica and 
Institut de Ciències del Cosmos (ICC), \\[1mm]
Universitat de Barcelona,\\[1mm] Martí i Franquès 1, 08028 Barcelona, Spain}
\emailAdd{jfumagalli@fqa.ub.edu}

\abstract{We question the existence of one-loop corrections to the large-scale power spectrum from small-scale modes in non-slow-roll dynamics which are not volume suppressed by the ratio of the short to long distance scales. One-loop contributions proportional to the long wavelength tree-level power spectrum, and not sharing this suppression, have
appeared in studies involving interactions singled out by the non-slow-roll dynamics.
In this context, we show the relevance of seemingly irrelevant interactions terms, such as the one provided by total derivative terms (boundary terms), and how they equally lead to non-volume suppressed contributions and exact cancellations. 
}

\begin{document}
\hfill{\flushright {}}
\maketitle
\newpage
\section{Introduction}

Primordial black holes \cite{Zeldovich:1967lct,Hawking:1971ei,Carr:1974nx,Meszaros:1974tb,Carr:1975qj,Chapline:1975ojl} are arguably the \textit{Occam's Razor} principle applied to Dark Matter \cite{Ivanov:1994pa,Garcia-Bellido:1996mdl}.
Tracing back their origin to an inflationary epoch requires dynamics beyond the vanilla slow-roll (SR) scenario which are able to enhance the amplitude of the primordial scalar fluctuations at short scales by several orders of magnitude compared to their value at CMB scales.  A standard way to achieve the desired magnification is via the introduction of a non-slow-roll phase along the inflationary dynamics, such as, for instance, ultra slow-roll (USR) \cite{Tsamis:2003px,Kinney:2005vj,Namjoo:2012aa,Germani:2017bcs,Motohashi:2017kbs,Ballesteros:2017fsr,Hertzberg:2017dkh} or constant-roll \cite{Motohashi:2014ppa, Inoue:2001zt,Tzirakis:2007bf}.  
How these small-scale enhanced scalar fluctuations influence large scales, for instance, the one probed by the CMB, has recently been the object of an intense debate \cite{Kristiano:2022maq,Riotto:2023hoz,Kristiano:2023scm,Riotto:2023gpm,Firouzjahi:2023aum,Firouzjahi:2023ahg,Franciolini:2023lgy,Tasinato:2023ukp,Cheng:2023ikq}. 
Computing one-loop contributions to the equal time correlators in this context leads to the appearance of corrections sharing the following form
\be\label{first}
\p_\R^{\rm 1-loop}(p)=c\,\, \p^{\rm tree}_\R(p) \int d\ln k \,\,\p^{\rm tree}_\R(k) +O\left(\frac{p^3}{k^3}\right),\qquad p\ll k,
\ee
where $\p_\R^{\rm tree / 1-loop}$ are the tree-level and one-loop dimensionless primordial power spectrum, $p$ and $k$ the momenta corresponding to the long and short scale modes respectively, and the integral domain is over the range of momenta spanned by the enhanced fluctuations. 
Corrections as the one in Eq. \eqref{first} has been first found in \cite{Kristiano:2022maq}, with 
$c\sim 10$ for a sharp transition from SR to USR and back to SR. Later, it has also been shown \cite{Riotto:2023hoz,Riotto:2023gpm,Firouzjahi:2023aum,Firouzjahi:2023ahg,Franciolini:2023lgy} that important suppression factors multiplying these contributions are present when considering realistic scenarios in which the transition from the ultra slow-roll to the slow-roll phase is smooth.

Taken at face value, Eq. \eqref{first} is telling us something, at the very least, counter-intuitive: 
 the relative 1-loop corrections to the tree-level power spectrum from small scales to large scales is independent of how widely separated the two scales are. 
 
Let us trace the source of this type of corrections.
Non-slow-roll scenarios, i.e. the ones under investigations here, are characterized by a sudden increase in absolute value of the so-called second slow-roll parameter $\eta$. That promptly calls attention to a particular subset of interaction terms present in the Hamiltonian of the scalar fluctuations. For instance, in \cite{Kristiano:2022maq}, that was the cubic interaction proportional to the time derivative of $\eta$. It is by considering one-loop diagrams built out of these selected terms alone that corrections as the one in \eqref{first} arise. 

In this work, we note that once additional diagrams ---which might seem irrelevant at first sight--- are included, corrections as the one highlighted in Eq. \eqref{first} precisely cancel. In short, we find the following result:
\be\label{result}
\mathrm{In \,\,Eq.\,\,}\eqref{first},\quad\,c=0.
\ee
We prove the previous statement in two different ways. The first is by considering the same interaction Hamiltonian taken into account in \cite{Kristiano:2022maq}, but now with the addition 
of a total time derivative term (boundary term) which is nevertheless present once the corresponding third order Lagrangian is written in this form.
Boundary terms may contribute in general to equal time correlators \cite{Seery:2006tq,Arroja:2011yj,Burrage:2011hd,Rigopoulos:2011eq,Agarwal:2013rva,Garcia-Saenz:2019njm}.
The intuition that boundary terms might be relevant in this context comes from the fact that the same terms do play a crucial role in computing contact diagrams such as the bispectrum \cite{Arroja:2011yj,Burrage:2011hd}. 

The second method consists of rewriting the action of the curvature perturbations in a form in which the previously introduced boundary term is inconsequential and the relevant bulk interactions are proportional to $\eta$ and not its first time derivative. By doing the computation in this way, namely with a different but equivalent set of interaction terms, we confirm the cancellation of the leading term
in Eq. \eqref{first}.  \\

\noindent\textbf{Note added.--}After the first version of this manuscript was released, the author in Ref. \cite{Firouzjahi:2023bkt} correctly remarked an important point. Although it is true that adding the boundary interaction considered in the present paper (see Eq. \eqref{Option1}) cancels the contribution highlighted in Eq. \eqref{first}, that interaction vertex alone does not grant the entire cancellation of all non-volume suppressed terms.
The original claim, i.e. the absence of all non-volume suppressed contributions induced by short modes to the long wavelength power spectrum, is valid only after considering our more complete analysis in Ref. \cite{Fumagalli:2024jzz}. The present study is thus restricted --and it has been modified accordingly throughout-- to showcase the importance of previously neglected boundary interactions when computing one-loop diagrams in transient non-slow-roll dynamics (and how one may equivalently take them into account via bulk interactions). 
Results of the current analysis also show, for the first time, the crucial role played in general by boundary terms in one-loop computations.

\section{Different methods to compute the relevant one-loop diagrams}
\subsection{Equivalent ways to express the third-order action}\label{Differentthird}

We consider standard single-field inflation and work in the gauge where the field fluctuations are set to zero, i.e. $\delta\phi=0$. This sets the time slicing.  The spatial part of the metric can be decomposed as $g_{ij}=a^2 e^{2\R} [e^{\gamma}]_{ij}$, where $a$ is the scale factor of a flat FLRW background, $\mathrm{tr}(\gamma)=0$ and $\R$ is the comoving curvature perturbation, the only degree of freedom needed here to describe the dynamics of the scalar fluctuations. %
One can further impose the gauge condition $\partial^{i}\gamma_{ij}=0$ to completely fix the gauge.\footnote{Note that this choice completely fixes the gauge, even non-linearly \cite{Maldacena:2002vr}, with the exception of finite volume effects that may be relevant only when dealing with infrared divergences—see \cite{Urakawa:2010it}. This aspect, as well as the question of whether fully fixing a gauge guarantees gauge independence—see \cite{Miao:2017feh,Glavan:2024elz}—goes beyond the scope of the present paper.}
The Lagrangian density for $\R$, here defined from the action as $\mathcal{S}\equiv \int d^3 x dt \mathcal{L}$,
is obtained after integrating the momentum and Hamiltonian constraints: 
\begin{align}\label{quadraticR}
\mathcal{L}^{(2)}&=\frac{1}{2} (\Mp^2 a^3 2\epsilon) \left[ \dot{\R}^2 - \frac{(\partial_i\R)^2}{a^2} \right], 
\end{align} 
where $\dot{}\equiv d/dt$ means derivative with respect to the cosmic time, $\epsilon \equiv -\dot{H}/{H^2}$ is the first slow-roll parameter, and $H \equiv \dot{a}/a$ the Hubble rate. We consider scenarios where the second slow-roll parameter, defined as
\be
\eta \equiv \frac{\dot{\epsilon}}{\epsilon \, H}, 
\ee
features a transient growth, i.e. the system experiences a non-slow phase where $|\eta| > 1$ in between two SR phases where $\epsilon, |\eta|  \ll 1$. Ultra slow-roll, characterized by a phase with $\eta=-6$, is a particular example of this type.   
A phase of large $\eta$ selects a subset of interaction terms from the cubic Lagrangian, i.e. it is natural to neglect terms with no instance of $\eta$ in this context. These latter give sub-leading contributions suppressed by the first slow-roll parameter which remains always small.
We first consider the third-order Lagrangian for $\R$ as derived in \cite{Maldacena:2002vr} once it is written to make explicit the size of the cubic interactions that start at second order in the slow-roll parameters. 
From there, we consider only terms with instances of the second slow-roll parameter $\eta$, i.e.

\begin{align}\label{L3mald}
\mathcal{L}^{(3)} =  \Mp^ 2\frac{a^3 \epsilon}{2}\dot{\eta} \zeta^2 \dot{\zeta} + \Mp^ 2\frac{d}{dt} \left[-\frac{a^3 \epsilon \eta}{2}\zeta^2\dot{\R} \,\right]  +  f(\R,\R')\frac{\delta \mathcal{L}^{(2)}}{\delta \R}.
\end{align}
To the original cubic Lagrangian present in \cite{Maldacena:2002vr} we have also included total derivative terms (boundary terms) which were not explicitly written there.\footnote{See e.g. \cite{Arroja:2011yj} for the full list of boundary terms corresponding to the third-order action in \cite{Maldacena:2002vr}.} In particular, we have considered a single total derivative term, while the ones not specified in Eq. \eqref{L3mald} are of two types: either they do not contain time derivative of the field $\R$ and so they do not contribute at any order to the correlation functions, or they give contributions to contact diagrams which are suppressed on super-Hubble scale\footnote{As discussed in Ref. \cite{Tada:2023rgp} and in our follow up work \cite{Fumagalli:2024jzz}, there is an additional cubic boundary term relevant for the full one-loop computation in the small momentum limit. In fact, as we stress therein, terms giving a suppressed contribution to contact diagrams on super-horizon scale are not necessary negligible when considering loop corrections.} ---see for instance \cite{Arroja:2011yj,Burrage:2011hd}. The third group of terms in Eq. \eqref{L3mald} 
labels terms proportional to the linear equation of motion (eom), i.e. the one satisfied by the free field, and we defined
\be\label{eqmot}
-\frac{1}{2 \Mp^ 2
}\frac{\delta \mathcal{L}^{(2)}}{\delta \R} \equiv \frac{d}{dt} (a^ 3 \epsilon \dot{\mathcal{\R}}) - a^ 3 \epsilon  \frac{\partial^2\R}{a^2}.
\ee
While terms proportional to the eom does not contribute to Feynman diagram at any order in perturbation theory, there is no a priori reason why the boundary term highlighted above should not contribute to equal time correlators. That is why we do not need, for the present purpose, to write explicitly the function $f$ in Eq. \eqref{L3mald} which is quadratic in $\R$.

Now, let us do a single manipulation (a would be ``integration by parts") on the first term in Eq. \eqref{L3mald} and rewrite it as follows
\begin{align}\nonumber
\Mp^ 2\frac{a^3 \epsilon}{2}\dot{\eta} \zeta^2 \dot{\zeta} &=  \Mp^ 2\frac{d}{d t}\left[ \frac{a^3 \epsilon \eta}{2}\R^2\dot{\R}  \right]-\Mp^ 2\eta\,\frac{d}{d t}\left[ \frac{a^3 \epsilon}{2}\R^2\dot{\R}  \right] \\[1mm]
&=\Mp^ 2\frac{d}{d t}\left[ \frac{a^3 \epsilon \eta}{2}\R^2\dot{\R}  \right] - \Mp^2 a^3\epsilon \eta \,\dot{\R}^2\R -\frac{\Mp^2}{2}\,\, a\epsilon \eta \,\R^2\partial^2\R + \frac{1}{4 
}\eta\,\R^2 \frac{\delta \mathcal{L}^{(2)}}{\delta \R},
\label{manipulation}
\end{align}
where we have expanded the time derivative in the second term on the first line and used Eq. \eqref{eqmot} to isolate terms proportional to the linear equation of motion. By substituting Eq. \eqref{manipulation} in Eq. \eqref{L3mald}, the total time derivative term cancels and the cubic Lagrangian becomes:
\be\label{mald2}
\mathcal{L}^{(3)} = - \Mp^2 a^3\epsilon \eta \,\dot{\R}^2\R -\frac{\Mp^2}{2}\,\, a\epsilon \eta \,\R^2\partial^2\R  + \tilde{f}(\R,\R')\frac{\delta \mathcal{L}^{(2)}}{\delta \R},
\ee
where we have reabsorbed in $\tilde{f}$ the term proportional to the equation of motion (last term in Eq. \eqref{manipulation} added to the last term in Eq. \eqref{L3mald}).\footnote{As a side note, this term actually cancels exactly a term proportional to the eom already present in the Lagrangian \eqref{L3mald}.}  

To compute correlators in transient non-slow roll scenarios, which are the focus of this work, we equivalently consider either the 
Lagrangian in Eq. \eqref{L3mald} or the 
one in Eq. \eqref{mald2}. We stress that including the total derivative term in Eq. \eqref{L3mald} (or its rewriting in Eq. \eqref{manipulation} leading to Eq. \eqref{mald2}) turned out to be crucial. 

\subsection{Equivalent Hamiltonians to compute one-loop diagrams } 

To compute correlators we switch to conformal time $d\tau = dt/a$. The Hamiltonian is obtained through a Legendre transform $\mathcal{H}=P\R'-a\mathcal{L}$, where $\mathcal{H}$ is the Hamiltonian density ($H=\int d^3 x \,\mathcal{H}$) and $P$ labels the conjugate momentum $P = \delta \mathcal{L}/\delta \R'$. 
For the present purpose, we are interested in the Hamiltonian up to cubic order in $\R$. Thus, we only need the linear relation between $\R'$ and $P$, i.e. $P_{\mathrm{lin}} = 2 a^2 \epsilon \R'$. 
The free Hamiltonian density is then\footnote{As it is customary in cosmological perturbation theory, we re-express the conjugate momentum in the Hamiltonian in terms of $\R'$. In the interaction picture the two are simply linearly related as discussed above.} $\mathcal{H}^{(2)}= P^2_{\mathrm{lin}}/(2 \Mp^2 a^2\epsilon) - a \mathcal{L}^{(2)}=\Mp^2\epsilon a^2(\R'^2+(\partial \R)^2)$, while the interaction Hamiltonian at cubic order is given by (see e.g. \cite{Chen:2010xka})
\be\label{cubic}
H_I^{(3)}=- \int d^3 x\, a \mathcal{L}^{(3)}.
\ee
We wish to compute 1PI one-loop diagrams following from the insertion of two cubic Hamiltonians.
In general, equal time correlators are computed with the $in-in$ formalism \cite{Schwinger:1960qe,Jordan:1986ug,Calzetta:1986ey}
\be\label{master}
\langle \R^{n}(t)\rangle=\langle0| \left[\bar{T}\left(e^{i\int_{-\infty(1+i\varepsilon)}^{\tau}d\tau' H_{I}(\tau')}\right)\right]\R_{I}^{n}(\tau)\left[T\left(e^{-i\int_{-\infty(1-i\varepsilon)}^{\tau}d\tau''\,H_{I}(\tau'')}\right)\right]|0\rangle \rangle,
\ee
where $T$ and $\bar{T}$ are the time and anti-time ordering operators, the subscript $I$ (that we neglect from now on) labels fields in the interaction picture, i.e. fields evolving with the linear equation of motions (with the free Hamiltonian), and the $\varepsilon$ prescription is there to project the adiabatic vacuum of the interacting theory into the vacuum of the free theory $|0\rangle$ in the infinite past. 
The perturbative expansion of Eq. \eqref{master} can be recast, in a compact form, in terms of a series of nested commutators --see, for instance, \cite{Chen:2010xka}. Although this rewriting looses the information on the UV regularization, the two formulations are equivalent as long as one is interested, as it is the case here, on the effects of interaction term which becomes relevant from a given preferred time.\footnote{UV divergences that one would usually find using the nested commutator form are not present in our computations simply because time integrals are cut at the time the interactions start to become relevant. 
By using the nested commutator form we will then re-find the same results present in the literature when considering the same interaction terms.}
For our purposes, we find more convenient to work in terms of the nested commutator form. This latter, for the one-loop contribution we are interested in, leads to
\begin{equation}\label{twonested}
\langle\hat{\R}_{\bp}(\tau)\hat{\R}_{\bp'}(\tau)\rangle_{\rm 1-loop}=-\int_{\tau_{\rm in}}^{\tau}d\tau_{1}\int_{\tau_{\rm in}}^{\tau_{1}}d\tau_{2}\langle [H^{(3)}(\tau_{2}),[H^{(3)}(\tau_{1}),\hat{\R}_{\bp}(\tau)\hat{\R}_{\bp'}(\tau)]]\rangle,
\end{equation}
where $\tau_{\rm in}$ labels a generic initial time selected by the interaction terms considered, and $\tau\rightarrow 0 $ is the time towards the end of inflation where correlators are evaluated.
  
Following our discussion in Sec. \ref{Differentthird},
where we have considered two rewriting of the same cubic Lagrangian in Eqs. \eqref{L3mald} and \eqref{mald2}, we compute one-loop correction to the two-point correlator $\langle \zeta_{\bp}\zeta_{\bp'}\rangle$ using two different although equivalent forms of the interaction Hamiltonian.
To derive them all we have to do is to insert Eqs. \eqref{L3mald} and \eqref{mald2} in \eqref{cubic}.

We thus proceed following the two equivalent methods summarized below and we obtain in both cases the same results.\footnote{There is at least a third method one may envisage. A standard way to remove boundary terms is by doing a field redefinition on $\R$, then compute correlation functions of the new variable and at the end relate the result to the correlation functions of the original $\R$ \cite{Maldacena:2002vr}. We plan to investigate this path, in this non-slow-roll context, in future works. Here we concentrate directly on the variable of interest, i.e. $\R$.} 
\begin{itemize}
    \item \hyperref[secmet1]{Method 1} 
    
    Insert in Eq. \eqref{twonested} two interaction Hamiltonian as derived from the Lagrangian \eqref{L3mald}: 
    \be\label{Option1} 
    H^{(3)}= \Mp^ 2\int d^3 x\left( -\frac{a^2 \epsilon}{2}\eta' \zeta^2 \zeta' + \frac{d}{d\tau} \left[\frac{a^2 \epsilon \eta}{2}\zeta^2\R'\right]\right).\ee
    \item \hyperref[secmet2]{Method 2} 
    
    Insert in Eq. \eqref{twonested} two interaction Hamiltonian as derived from the Lagrangian \eqref{mald2}:
    \be \label{Method2}H^{(3)}=\Mp^ 2\int d^3 x \left( a^2\epsilon \eta  (\R')^2\R + \frac{1}{2} a^2 \epsilon\eta \R^2 \partial^2\R\right). \ee
\end{itemize}
In both expressions above, we have neglected terms proportional to the linear equation of motion, as they vanish identically when the Hamiltonian is inserted into Eqs. \eqref{master} and \eqref{twonested}, since the fields in the interaction picture satisfy the linear equation of motion. Thus, we emphasize that \eqref{Option1} and \eqref{Method2} are merely different representations of the same Hamiltonian (with the same canonical variables), which differ up to terms proportional to the linear equation of motion irrelevant in the operator formalism used in this work.

\section{One-loop corrections in non-slow-roll dynamics}

\subsection{Conventions and approximations}
Let us list the few ingredients and approximations needed for our purposes.
In momentum space\footnote{We use the Fourier transform convention $f(\bx)=(2\pi)^{-3}\int d\bk\, e^{-i\bk\cdot\bx}f(\bk)$.}, canonical quantization of the free fields
leads to
\be\label{interactionfield}
\Rh_{\bp}(\tau) =\R_{p}(\tau) \hat{a}(\bp) + \R^*_{p}(\tau) \hat{a}^{\dagger}(-\bp),
\ee
with annihilation/creation operators satisfying standard commutation relations
\be\label{quantiz}
[\hat{a}(\bp),\hat{a}^{\dagger}(\bp')]= (2\pi)^3 \delta(\bp -\bp'),
\ee
and $\R_p(\tau)$ labelling the mode functions of the free fields, solutions of the linear equations of motion.
From the two previous relations, we write explicitly the following equal time commutators:
\be\label{firstcommutator}
\left[\Rh_{\bk}(\tau),\Rh_{\bk'}(\tau)\right]\equiv(2\pi)^3\delta(\bk+\bk') 2i\,\mathrm{Im}(|\R_k|^2) = 0,
\ee
and\footnote{Eq. \eqref{building} trivially leads to $[\Rh(\bx,\tau),\hat{P}_{\rm lin}(\bx',\tau)]=i \delta(\bx-\bx')$ where $\Rh$ and $\hat{P}_{\rm lin}=2a^2\epsilon \Rh'$ are the linear field and conjugate momentum.}
\be\label{building}
\left[\Rh_{\bk}(\tau),\Rh'_{\bk'}(\tau)\right]\equiv(2\pi)^3\delta(\bk+\bk')2i\, \mathrm{Im}(\R_k\R_k^{*'})=(2\pi)^3\delta(\bk+\bk')\, \frac{i}{2 a^2\Mp^ 2\epsilon}.
\ee
Fields commute at equal time and last equality comes from the Wronskian normalization imposed over the mode functions to have standard commutation relations as in Eq. \eqref{quantiz}.

We consider a non-slow-roll phase in between two standard slow-roll evolutions.
A simple way to model this scenario, while working directly at the level of the effective field theory of the perturbations, is by using the following top-hat profile for $\eta$:
\be
\eta = \tilde{\eta} +  \tilde{\tilde{\eta}},\,\,\,\mathrm{with}\,\,\,\,
\tilde{\eta} = -\Delta\eta \,\theta(\tau-\taue),\quad \mathrm{and}\,\,\,\,\,\tilde{\tilde{\eta}} = \Delta\eta \,\theta(\tau-\taus),
\ee
$\taus$ and $\taue$ label respectively the start and the end of the non-slow-roll phase. 
We stress that our conclusions are independent of the sharpness of the transition, and $\theta$, here labelling the Heaviside function, can also be replaced by smoother alternatives. Within our definitions, what it is usually called ultra slow-roll corresponds to $\Delta\eta = -6$.

$\eta'$ peaks at $\tau_s$ and $\tau_e$. For simplicity and to capture leading effects when considering interaction terms proportional to $\eta'$ in the correlators, we only consider the peak at $\tau_e$. At $\tau_s$ all mode functions are still following their slow-roll, not enhanced, evolution and we thus disregard, for simplicity, these contributions. 
Thus, we use the following approximation
\begin{align}\label{approximation1}
\int^{\tau_1} d\tau_2 \, \eta'(\tau_2) f(\tau_2) \simeq \int ^{\tau_1} d\tau_2\,\tilde{\eta}'(\tau_2)f(\tau_2)=-\Delta\eta\,\theta(\tau_1-\tau_e)f(\tau_e)= \tilde{\eta}(\tau_1)f(\tau_e).
\end{align}
with $f(\tau)$ a continuous function at $\tau_e$.\footnote{Although re-written according to our taste, this discussion leads exactly to the same approximation present, for instance, in \cite{Kristiano:2022maq}.} 
For simplicity, in this study we always work under the approximation $ p \ll k$ where $\bk$ label modes exiting the horizon around the times the ultra slow-roll dynamics takes place while $\bp$ labels momenta associated to the long wave modes. 
We consider the wavelength associated to $p$ well outside the horizon before reaching the non-slow-roll region. $\tau_i \in [\tau_s,\tau_e]$ denotes times during the non-slow-roll phase and $\tau$ labels the end of inflation.
We can thus use the following approximation for the unequal time commutator between $\Rh$ and its first derivative:
\be\label{approxcomm2}
[\Rh_{\bp}(\tau),\Rh'_{\bk}(\tau_i)]=(2\pi)^3\delta(\bp+\bk) \frac{i}{2 a^2\Mp^2 \epsilon} + O(p\tau, p\tau_i).
\ee
As will be evident from the computation in Sec. \ref{secmet1}, the momentum independence of the right-hand side of the above commutator at leading order leads to the emergence of corrections in the form shown in Eq. \eqref{first}.

Further, let us consider the unequal time commutator between two fields. That can be written as
\be\label{approxcomm}
[\Rh_{\bk}(\tau_i),\Rh_{\bp}(\tau)]= (2\pi)^3\delta(\bp +\bk)\frac{i}{2 a^2\Mp^2 \epsilon} g_p(\tau,\tau_i) ,
\ee
where $g_p$ denotes the Green's function associated with the linear equation of motions.
In the super-horizon limit, i.e. $p\tau_i,p\tau \ll 1$, $g_p\propto \tau_i$. Namely, the Green's functions loose their momentum dependence at leading order. Thus, one-loop diagrams proportional to the commutator in Eq. \eqref{approxcomm}, lead equivalently to non-volume suppressed terms--see \cite{Firouzjahi:2023bkt,Fumagalli:2024jzz}.

As discussed in Sec. \ref{secmet1}, the first operator in Eq. \eqref{Option1}, i.e. the one first considered in \cite{Kristiano:2022maq}, does not give a contribution proportional to the Green's function.
The scope of the present analysis is restricted to emphasize the role of total derivative terms in relation to contributions from bulk operators.
We thus consider, from the new interaction included here, i.e. the total derivative term in Eq. \eqref{Option1}, only contributions not proportional to the Green's function. The same applied to the equivalent computation performed in Sec. \ref{secmet2}. We will see that up to these terms, the boundary operator considered exactly cancels the contribution from the bulk operator in Eq. \eqref{Option1}. The full computations, also including relevant quartic interactions, is then postponed to the follow up analysis in Ref. \cite{Fumagalli:2024jzz}. 

To our taste, in the context under investigation here, using the nested commutator form in Eq. \eqref{twonested}, together with the compact notation defined in the next section --see Eq. \eqref{compact}-- allow us to considerably simplify computations. As it will be shown, after unfolding the nested commutators, one can use directly the commutation relations between fields and fields derivative \eqref{firstcommutator}-\eqref{building} and \eqref{approxcomm2} (\eqref{approxcomm}).  
These small set of relations is all we need to proceed with our computation.

We are ultimately interested in computing the dimensionless power spectrum $\p_\R$ defined as
\be\label{dimdef}
\langle \hat{\R}_{\bp}(\tau) \hat{\R}_{\bp'}(\tau)\rangle =  \delta(\bp + \bp') (2\pi)^3 \frac{2\pi^2}{p^3}\mathcal{P}_\R(p,\tau).
\ee
$\p_\R$ is computed perturbatively, i.e.
\be
\p_\R = \p_\R^{\mathrm{tree}} + \p_\R^{\mathrm{1-loop}}+ ...\,,
\ee
where the tree level power spectrum $\p_\R^{\mathrm{tree}}$, obtained by taking the correlators of linear fields, is given by\footnote{Standard Wick contraction between two linear modes gives $ \wick{\c1{\hat{\R}_{\bp}}(\tau) \c1{\hat{\R}_{\bp'}}(\tau)}  =  \delta(\bp + \bp') (2\pi)^3 |\R_p|^ 2$.}
\be
\mathcal{P}_\R^{\mathrm{tree}}= \frac{p^3}{2\pi^2}|\R_p|^2,
\ee
while the one-loop corrections $\p_\R^{\mathrm{1-loop}}$, computed in the next sections, are given by considering, in Eq. \eqref{dimdef}, the correlator in Eq. \eqref{twonested}.

\subsection{Method 1}\label{secmet1}

Let us compute \eqref{twonested} with the interaction Hamiltonian \eqref{Option1} divided in the two sub-pieces
\begin{align}\label{Option1b} \nonumber
    H^{(3)}&=\,\, H^{(3)}_a +H^{(3)}_b \,\\ 
    H^{(3)}_a &\equiv \Mp^ 2\int d^3 x\left( -\frac{a^2 \epsilon}{2}\eta' \zeta^2 \zeta'\right),\qquad 
    H^{(3)}_b \equiv \Mp^ 2\int d^3x\left(\frac{d}{d\tau} \left[\frac{a^2 \epsilon \eta}{2}\zeta^2\R'\right]\right).
\end{align}

\subsubsection{Contribution from bulk interactions}
We start by considering only instances of $H^{(3)}_a$. That allows us to recover the result first found in \cite{Kristiano:2022maq} by means of the nested commutator form.
From \eqref{twonested} we can write
\begin{align} \nonumber
\langle \Rh_{\bp}\Rh_{\bp'}\rangle_{[a,\,a]}\,\equiv\,&-\int^{\tau}d\tau_{1}\int^{\tau_{1}}d\tau_{2}\langle [H^{(3)}_{a}(\tau_{2}),[H^{(3)}_{a}(\tau_{1}),\hat{\R}_{\bp}(\tau)\hat{\R}_{\bp'}(\tau)]]\rangle\\ \label{aa}
\equiv \,& \int^\tau d\tau_{1} \eta'(\tau_1)\int^{\tau_{1}}d\tau_{2} \eta'(\tau_2) \langle\hat{f}_{a}(\tau_2,\tau_1)\rangle,
\end{align}
where 
\be\label{deff}
\hat{f}_{a}(\tau_2,\tau_1) \equiv -\frac{\Mp^ 4}{4} a^2(\tau_1)\epsilon(\tau_1)a^2(\tau_2)\epsilon(\tau_2)\int d\bK \left[ \Rh_{\bk_1}\Rh_{\bk_2}\Rh'_{\bk_3}\Huge|_{\tau_2},  \left[ \Rh_{\bk_4}\Rh_{\bk_5}\Rh'_{\bk_6}\Huge|_{\tau_1}, \Rh_{\bp}\Rh_{\bp'}  \right] \right]
\ee
and Fourier transforming has left us with 
\be\label{bigK}
\int d\bK \equiv \prod_{i=1}^{3}\left[\int \frac{d\bk_{1,i}}{(2\pi)^ 3}\right](2\pi)^ 3\delta\left(\sum_{i=1}^{3} \bk_{1,i}\right)   \prod_{i=1}^{3}\left[\int \frac{d\bk_{2,i}}{(2\pi)^ 3}\right] (2\pi)^ 3 \delta\left(\sum_{i=1}^{3} \bk_{2,i}\right).
\ee
Before unfolding the nested commutator let us apply multiple times the approximation in Eq. \eqref{approximation1} to Eq. \eqref{aa}:
\begin{align}\nonumber
\langle \Rh_{\bp}\Rh_{\bp'}\rangle_{a}\,&\simeq\, \int^\tau d\tau_{1} \eta'(\tau_1)\int^{\tau_{1}}d\tau_{2} \, \tilde{\eta}'(\tau_2) \langle\hat{f}_{a}(\tau_e,\tau_1)\rangle
 =  \int^\tau d\tau_{1} \eta'(\tau_1)\tilde{\eta}(\tau_1) \langle\hat{f}_{a}(\tau_e,\tau_1)\rangle \\ \label{resultaa}
 &\simeq \int^{\tau} \frac{(\tilde{\eta}\tilde{\eta})'}{2}  \langle\hat{f}_{a}(\tau_e,\tau_e)\rangle = \frac{(\Delta\eta)^ 2}{2} \langle\hat{f}_{a}(\tau_e,\tau_e)\rangle.
\end{align}
Let us now turn our attention to the nested commutator inside $\langle\hat{f}_a(\tau_e,\tau_e)\rangle$. To simplify notation and focus on the commuting structure, we label with the same index operators coming from the same vertex, i.e.
\be\label{compact}
\left[ \Rh_{\bk_{1,1}}\Rh_{\bk_{1,2}}\Rh'_{\bk_{1,3}}\Huge|_{\tau_2},  \left[ \Rh_{\bk_{2,1}}\Rh_{\bk_{2,2}}\Rh'_{\bk_{2,3}}\Huge|_{\tau_1}, \Rh_{\bp}\Rh_{\bp'}\Huge|_{\tau} \right] \right]\equiv \left[ \Rh^ 2_{2}\Rh'_{2},  \left[ \Rh_1^2\Rh'_1, \Rh_{\bp}\R_{\bp'} \right] \right]. 
\ee
Note that, after doing all contractions and integrating over the Dirac deltas, the overall momentum conservation allows us to simply replace $k$ and $|\bp-\bk|$ to the momenta of the mode functions corresponding to internal contractions (contractions of $\Rh$'s corresponding to different vertexes) and $p$ to the momenta of the mode functions corresponding to contractions with external legs, all that by leaving behind an overall integration $\int d \bk$.
From the approximation \eqref{resultaa} and using notation \eqref{compact}, what we wish to compute takes the form
\begin{align}\label{ff}
\langle \Rh_{\bp}\Rh_{\bp'}\rangle_{[a,\,a]} =&-|\Delta\eta|^ 2 \frac{\Mp^ 4 a^ 4 \epsilon^ 2\Huge|_{\tau_e}
}{8} \int d\bK \langle\left[ \Rh^ 2_{2}\Rh'_{2},  \left[ \Rh_1^2\Rh'_1, \Rh_{\bp}\Rh_{\bp'} \right]\right]\rangle\Bigg|_{\tau_1=\tau_2=\tau_e}.
\end{align} 
Note that, since we are considering two insertions of the bulk operator $H^{(3)}_a\propto \eta'$ in Eq. \eqref{Option1b}, which is a Dirac delta in our approximation, the two interactions terms turned out to be evaluated at the same time in Eq. \eqref{ff}. Let us then proceed to expand the nested commutator in Eq. \eqref{ff}.

The would be contribution proportional to the Green's function, i.e. proportional to $[\Rh_1,\Rh_{\bp}]$---see Eq. \eqref{approxcomm}, takes the form
\begin{align}
\langle \Rh_{\bp}\Rh_{\bp'}\rangle_{[a,\,a]} \supset [\Rh_1,\Rh_{\bp}]\cdot [\Rh'_2\Rh_2\Rh_2,\Rh'_1\Rh_1\Rh_{\bp}] \simeq 0,
\end{align}
and --in this case-- vanishes because the second commutator on the right-hand side is, within our approximation, the equal-time commutator of the same operator.\footnote{This is true if one is legitimate to consider $\Rh_{\bp}$ as evaluated at $\tau_e$ instead of at the end of inflation $\tau$. Let us explain why this is the case within our approximation. First, note that at this stage, only taking $\Rh_{\bp}$ outside the commutator may give a non-volume suppressed contribution. Next, once $\Rh_{\bp}$ is taken outside the commutator, its time dependence no longer plays a significant role. In the limit $p\tau, p\tau_e \ll 1$, one obtains $\langle \Rh_2 \Rh_{\bp} \rangle \propto \mathrm{Re}\left(\zeta_p(\tau_e) \zeta^*_p(\tau)\right) \simeq |\R_p|^2$. The last expression can be equivalently considered as evaluated at either $\tau_e$ or $\tau$.
}
That simplifies considerably the computation which is then almost trivialized in the few lines below:
\begin{align}\nonumber
\left[ \Rh^ 2_{2}\Rh'_{2},  \left[ \Rh_1^2\Rh'_1, \Rh_{\bp}^ 2 \right]\right]=&  2\left[\Rh'_1,\Rh_{\bp}\right] \cdot  \left[ \Rh^ 2_{2}\Rh'_{2},  \Rh_1^2 \Rh_{\bp}\right]\\ \nonumber
=& 2\left[\Rh'_1,\Rh_{\bp}\right] \cdot \Rh^ 2_{2}\left(  \left[ \Rh'_{2},  \Rh_1^2 \right] \Rh_{\bp}+   \Rh_1^2\left[ \Rh'_{2},  \Rh_{\bp}\right]\right)\\ \label{unfoldcomm} 
=&2\left[\Rh'_1,\Rh_{\bp}\right] \cdot \left(  2\left[ \Rh'_{2},  \Rh_1 \right] \Rh^ 2_{2}\Rh_1\Rh_{\bp}+   \left[ \Rh'_{2},  \Rh_{\bp}\right] \Rh^ 2_{2}\Rh_1^2\right).
\end{align}
We now replace each building block commutator with Eqs. \eqref{building}-\eqref{approxcomm2} which cancels the prefactor $\Mp^4 a^4\epsilon^2$ in \eqref{ff}, and we then Wick contract operators left outside the commutators. That promptly leads to 
\be\label{finalaa1}
\langle \Rh_{\bp}\Rh_{\bp'}\rangle_{[a,\,a]}  =\delta(\bp+\bp') \frac{|\Delta\eta|^2}{4}\int d\bk \left(|\zeta_p|^ 2|\zeta_{k}|^ 2 + \frac{1}{2}|\zeta_k|^2 |\zeta_{|\bp-\bk|}|^ 2\right).
\ee
The second term is a convolution of the modes running in the loop. Once going to the dimensionless power spectrum \eqref{dimdef}, that correction is volume suppressed, i.e. scale as the ratio of the short to the long scales cube:
\be\label{volume}
\int d\ln k \, k^ 3 |\zeta_k'|^ 2 |\zeta_{|\bp-\bk|}|^ 2 \propto \frac{1}{k^ 3} \,\,\implies \,\,\,\p^{\rm 1-loop}_\zeta(p) \propto \left(\frac{p}{k}\right)^ 3\cong 0.
\ee
We mention this explicitly only once here and neglect this type of contributions in the following sections.\footnote{One can check that, by using the approximation \eqref{approxcomm}, we had already neglected volume suppressed terms proportional to convolution of first derivatives of the mode functions with momenta $k$ and $\q$.} 

We use the symbol $\cong$ to identify quantities which are \textit{equal up to volume suppressed terms} and later also to identify quantities up to terms proportional to the Green's function/commutator in Eq. \eqref{approxcomm}.

Considering the first and dominant term in Eq. \eqref{finalaa1} and using the definition \eqref{dimdef} we get, with obvious notation,
\be\label{fin}
\p^{\rm 1-loop}_{\R,\,[a,a]}(p) \cong \frac{|\Delta\eta|^2}{4} \p_\R^{\rm tree}(p) \int \frac{d\bk}{(2\pi)^3} |\zeta_{k}(\taue)|^ 2 = \frac{|\Delta\eta|^2}{4} \p_\R^{\rm tree}(p) \int d\ln k\, \p^{\rm tree}_\zeta (k,\taue)  .
\ee
Last expression has the form of the first term in Eq. \eqref{first} with $c= |\Delta\eta|^2/4$. This contribution, first found in \cite{Kristiano:2022maq}, is in principle logarithmically divergent and would require proper regularization.
\footnote{Therein, the integral is evaluated within a window that encompasses only the enhanced scales, in order to capture their contribution to the one-loop correction on the long-wavelength modes.} We are now going to show that diagrams constructed from the total derivative term in Eq. \eqref{Option1b} yield a contribution equal but with opposite sign to the one in Eq. \eqref{fin}.
\subsubsection{Contributions from boundary terms}
Let us now compute contributions from boundary terms. By inserting two instances of $H_b$ defined in Eq. \eqref{Option1b} into Eq. \eqref{twonested} we get
\begin{align}\nonumber
\langle \Rh_{\bp}\Rh_{\bp'}\rangle_{[b,\,b]}\,\equiv\,&-\int^{\tau}d\tau_{1}\int^{\tau_{1}}d\tau_{2}\langle [H^{(3)}_{b}(\tau_{2}),[H^{(3)}_{b}(\tau_{1}),\hat{\R}_{\bp}(\tau)\hat{\R}_{\bp'}(\tau)]]\rangle\\ \label{Optionbnotation}
\equiv&  \int^ {\tau} d\tau_1 \int^ {\tau_1} d \tau_2 \langle \hat{f}_{b}(\tau_2,\tau_1)\rangle
\end{align}
with
\begin{align}
\hat{f}_{b}(\tau_2,\tau_1) \equiv  -\frac{\Mp^4}{4} \int d \bK\left[ \left(a^2 \epsilon \eta\Rh_{\bk_{2,1}}\Rh_{\bk_{2,2}}\Rh'_{\bk_{2,3}}\right)'\Big|_{\tau_2} ,\left[ \left(a^2 \epsilon \eta\Rh_{\bk_{1,1}}\Rh_{\bk_{1,2}}\Rh'_{\bk_{1,3}}\right)'\Big|_{\tau_1} ,\Rh_{\bp}\Rh_{\bp'}\right]\right],
\end{align}
where $d\bK$ was defined in Eq. \eqref{bigK}.
Integration over $\tau_2$ can be performed straightforwardly. Then we expand the derivative in the first nested commutator and obtain
\begin{align}
\label{bbpartial}
\langle \Rh_{\bp}\Rh_{\bp'}\rangle_{[b,\,b]}\,=\,
& -\frac{\Mp^ 4}{4}  \int d \bK \int^ {\tau} d\tau_1  \langle \left[ a^2 \epsilon \eta\Rh^2\Rh' ,\left[ \eta(a^2 \epsilon\Rh')'\Rh^2 + \eta' a^2 \epsilon \Rh^2\Rh' +  2 a^2 \epsilon(\Rh')^ 2 \Rh ,\Rh_{\bp}^2\right]\right]\rangle,
\end{align}
where all operators in the nested commutator are evaluated at the same time $\tau_1$.
By using the linear equations of motion to rewrite the first term in the nested commutator in Eq. \eqref{bbpartial}, i.e. $(a^2 \epsilon\Rh')'= a^ 2\epsilon\partial^ 2\Rh$, one notices that this term gives contribution proportional to the commutator in Eq. \eqref{approxcomm}, that we disregard for the reasons discussed below Eq. \eqref{approxcomm}.
Applying, as in the previous section, the same approximations \eqref{approximation1} to the second term in the nested commutator in Eq. \eqref{bbpartial} we get
\be
-\frac{\Mp^ 4}{4} \int d \bK \int^ {\tau} \frac{(\eta\eta)'}{2} a^4 \epsilon^2 \langle\left[\Rh_2\Rh_2\Rh_2',\left[  \Rh_1\Rh_1\Rh_1' ,\Rh_{\bp}\Rh_{\bp'}\right]\right] \rangle\simeq -\frac{|\Delta\eta|^2}{2}\langle\hat{f}_a(\tau_e,\tau_e)\rangle = - \langle \Rh_{\bp} \Rh_{\bp'}\rangle_{[a,a]},
\ee
where $\hat{f}_a$ was defined in Eq. \eqref{deff}. By adding the third term in the nested commutator in Eq. \eqref{bbpartial} we can rewrite this all contribution as
\be\label{bbpartial22}
\langle \Rh_{\bp} \Rh_{\bp'}\rangle_{[b,b]} = - \langle \Rh_{\bp} \Rh_{\bp'}\rangle_{[a,a]}  -\Mp^ 4\frac{|\Delta\eta|^ 2}{2}  \int d \bK \int^ {\taue}_{\taus} d\tau_1 a^4 \epsilon^2  \langle \left[ \Rh_2\Rh_2\Rh_2' ,\left[  \Rh'_1 \Rh'_1
\Rh_1,\Rh_{\bp}\Rh_{\bp'}\right]\right] \rangle.
\ee
The expression above already suggests that the contribution from the boundary interaction is not irrelevant, i.e. it provides a term equal but with opposite sign with respect to the bulk operator considered in the previous section.
By proceeding analogously one can show that 
\begin{align}\label{cross}
\langle \Rh_{\bp}\Rh_{\bp'}\rangle_{[a,\,b]}\, \cong -\langle \Rh_{\bp}\Rh_{\bp'}\rangle_{[b,\,a]} \cong - \langle \Rh_{\bp}\Rh_{\bp'}\rangle_{[a,\,a]},
\end{align}
where, with the same notation as in Eq. \eqref{Optionbnotation}, $\langle \Rh_{\bp}\Rh_{\bp'}\rangle_{[a,\,b]}$ refers to Eq. \eqref{twonested} with the insertions $H^{(3)}_{a}(\tau_{2})$ and $H^{(3)}_{b}(\tau_{1})$, while $\langle \Rh_{\bp}\Rh_{\bp'}\rangle_{[b,\,a]}$ refers to Eq. \eqref{twonested} with the insertions $H^{(3)}_{b}(\tau_{2})$ and $H^{(3)}_{a}(\tau_{1})$.
From Eq. \eqref{cross} we thus note that the total contribution from boundary terms is just given by $\langle \Rh_{\bp}\Rh_{\bp'}\rangle_{[b,\,b]}$.  

From \eqref{bbpartial22} we proceed as in Eq. \eqref{unfoldcomm} (also symmetrizing not commuting operators\footnote{Operators which do not commute at equal time have always to be thought as in their symmetrized version, i.e. $(\Rh'_1)^2\Rh_1=1/2((\Rh'_1)^2\Rh_1 + \Rh_1(\Rh'_1)^2)$, we did not write that explicitly at each step just to leave the notation lighter, but we performed the full computations with the symmetrized operators.}).
After tedious manipulations we get
\begin{align}\nonumber
 \left[  \Rh'_1 \Rh'_1
\Rh_1,\Rh_{\bp}\Rh_{\bp'}\right]&= 4\left[  \Rh'_1,\Rh_{\bp'} \right]\left(2\Rh_{\bp}\Rh_2\Rh_1\Rh_2'\left[  \Rh_2,\Rh_1' \right] + \Rh_2^ 2 \Rh_1\Rh_1' \left[  \Rh_2',\Rh_{\bp}\right] + \Rh_2^ 2 \Rh_{\bp}\Rh_1' \left[  \Rh_2',\Rh_1  \right] \right. \\
 &\quad \left. \vphantom{} \,\,+  
 2\Rh'_{2}\Rh_1\Rh_{\bp}\Rh_2\left[  \Rh_2,\Rh_1' \right] + \Rh_1 \Rh_1'\Rh_2^ 2 \left[  \Rh_2',\Rh_{\bp}\right] + \Rh_{\bp}\Rh_1'\Rh_2^ 2 \left[  \Rh_2',\Rh_1  \right]
\right),
\end{align}
where again we have disregarded terms proportional to the commutator in Eq. \eqref{approxcomm}.

We use the approximation in Eq. \eqref{approxcomm2} to evaluate the commutator outside the parenthesis. Further, all commutators inside parenthesis are evaluated at equal time so that we can use Eq. \eqref{building}. All that cancel the $4 a^2 \epsilon^2$ prefactor. Wick contracting operators outside commutators, and using the power spectrum definition in Eq. \eqref{dimdef} lead to (still neglecting sub-leading terms proportional to $\R_p'$)
\begin{align}\nonumber
\p^{\rm 1-loop}_{\R,\,[b,b]} + \p^{\rm 1-loop}_{\R,\,[a,a]} \cong&
 -\frac{|\Delta\eta|^ 2}{2} \frac{p^ 3}{2\pi^ 2}\int \frac{d \bk}{(2\pi)^ 3} \int d\tau_1 \left( 
|\R_p|^2(|\R_k|^ 2)' - |\R_p|^2(|\R_k|^ 2)'  -|\R_k|^2(|\R_{|\bp-\bk|}|^2)' 
\right)\\ 
\cong &\,\, 0.
\end{align}
Note that terms proportional to the long-wave power spectrum, namely the first two terms, exactly cancel. The last term just gives a volume suppressed contribution, i.e. scales as $(p/k)^3$. As already discussed, the symbol $\cong$ identifies quantities up to volume suppressed terms and up to term proportional to the Green's function in Eq. \eqref{approxcomm}. The latter type of contribution has been neglected from diagrams associated to the boundary interaction.

\subsection{Method 2}\label{secmet2}
We now carry out an equivalent derivations which does not require the introduction of the boundary terms. In a sense, this is a more orthodox way of proceeding. We compute the correlator in Eq. \eqref{twonested} with the interaction Hamiltonian in Eq. \eqref{Method2}. As before, it is useful to divide the Hamiltonian in two sub-pieces:
\begin{align} \nonumber
H^{(3)}&=H_c^{(3)}+H_d^{(3)} \\ \label{defcccd}
H_c^{(3)} &= \Mp^ 2\int d^3 x  a^2\epsilon \eta  (\R')^2\R,\quad H_d^{(3)} =\frac{1}{2} \Mp^ 2 \int d^3 x  a^2 \epsilon\eta \R^2 \partial^2\R.
\end{align}
From \eqref{twonested} we have two non-trivial contributions:
\begin{align}\label{cc}
\langle \Rh_{\bp}\Rh_{\bp'}\rangle_{[c,\,c]}\,\equiv\,-\int^{\tau}d\tau_{1}\int^{\tau_{1}}d\tau_{2}\langle [H^{(3)}_{c}(\tau_{2}),[H^{(3)}_{c}(\tau_{1}),\hat{\R}_{\bp}(\tau)\hat{\R}_{\bp'}(\tau)]]\rangle,
\end{align}
and
\begin{align}\label{dc}
\langle \Rh_{\bp}\Rh_{\bp'}\rangle_{[d,\,c]}\,\equiv\,-\int^{\tau}d\tau_{1}\int^{\tau_{1}}d\tau_{2}\langle [H^{(3)}_{d}(\tau_{2}),[H^{(3)}_{c}(\tau_{1}),\hat{\R}_{\bp}(\tau)\hat{\R}_{\bp'}(\tau)]]\rangle.
\end{align}
Terms where the Hamiltonian in the first nested commutator is equal to \( H_d^{(3)} \), such as \( \langle \Rh_{\bp}\Rh_{\bp'}\rangle_{[c,\,d]} \) and \( \langle \Rh_{\bp}\Rh_{\bp'}\rangle_{[d,\,d]} \), give contributions proportional to the commutator in Eq.~\eqref{approxcomm}, i.e., \( \propto [\Rh_1, \Rh_{\bp}] \). This is simply due to the fact that \( H_d^{(3)} \) does not contain time derivative operators. As before, we do not include these contributions here.

\subsubsection{[c-c] contribution}
Let us consider two instances of $H_c$, as defined in \eqref{defcccd}, into Eq. \eqref{twonested}:
\begin{align} \nonumber
\langle \Rh_{\bp}\Rh_{\bp'}\rangle_{[c,\,c]} &= -\Mp^ 4|\Delta\eta|^2\int^{\taue}_{\taus} d\tau_1 (a^2\epsilon)|_{\tau_1} \int^{\tau_1}_{\tau_s} d\tau_2 (a^2\epsilon)|_{\tau_2}\\ \label{cc22}
&\quad \times\int d\bK \langle [(\Rh'_{\bk_1}\Rh'_{\bk_2}\Rh_{\bk_3})|_{\tau_2},[(\Rh'_{\bk_4}\Rh'_{\bk_5}\Rh_{\bk_6})|_{\tau_1},\Rh_{\bp}\Rh_{\bp'}]] \rangle.
\end{align}
We start by unfolding the first nested commutator. Using the same compact notation defined in Eq. \eqref{compact}, that becomes  
\begin{align}\nonumber
[\Rh'_1\Rh'_1\Rh_1,\Rh_{\bp}\Rh_{\bp'}]=\,&[\Rh'_1,\Rh_{\bp'}](\Rh_1'\Rh_{\bp}\Rh_1 + \Rh_{\bp}\Rh_1'\Rh_1 + \Rh_1\Rh_1'\Rh_{\bp}+ \Rh_1\Rh_{\bp}\Rh_1') \\ \nonumber
=\,&[\Rh'_1,\Rh_{\bp'}](2\Rh_1'\Rh_{\bp}\Rh_1 + [\Rh_{\bp},\Rh_1']\Rh_1+2\Rh_1\Rh_1'\Rh_{\bp}+ \Rh_1[\Rh_{\bp},\Rh_1'])\\ \label{firstun}
=\,&[\Rh'_1,\Rh_{\bp'}](4\Rh_1'\Rh_1\Rh_{\bp} + 2[\Rh_1,\Rh_1']\Rh_{\bp}+ 2\Rh_1[\Rh_{\bp},\Rh_1']).
\end{align}
In the second line we have added and subtracted terms equal to the first and third one in parenthesis on the first line. On the third line we added and subtracted a term equal to the first in parenthesis on the second line and used the fact that we are neglecting terms proportional to $[\Rh_{\bp},\Rh_1]$. Terms proportional to commutators in the parenthesis of the final expression just lead to tadpole diagrams, namely they force contractions between two legs from the same vertexes once embedded in the full expression \eqref{cc22}. As before we focus on 1PI diagrams and so we disregard these terms. 
Inserting Eq. \eqref{firstun} in the full nested commutator in \eqref{cc22}, and after similar manipulations as the one just discussed, we get
\be\label{fullnestedcc}
[\Rh'_2\Rh'_2\Rh_2,[\Rh'_1\Rh'_1\Rh_1,\Rh_{\bp}\Rh_{\bp'}]]= 8[\Rh'_1,\Rh_{\bp'}]\left( [\Rh'_2,\Rh'_1]\Rh_2'\Rh_2\Rh_1\Rh_{\bp} + [\Rh'_2,\Rh_1]\Rh_1'\Rh_2'\Rh_2\Rh_{\bp}+ [\Rh'_2,\Rh_{\bp}]\Rh_1'\Rh_1\Rh_2'\Rh_2\right).
\ee
Last term in the previous equation just lead to a volume suppressed contribution, see \eqref{volume}. 
Let us then consider only the first two terms in \eqref{fullnestedcc}. After the usual Wick contractions we have that the one-loop correction to the power spectrum from these terms can be written as:\footnote{Eq. \eqref{ccpartial} coincides with the leading order result found in Ref. \cite{Firouzjahi:2023aum} (see Appendix A, Eq. (97) of version 1), with the exception that we found a crucial time switch $\tau_1 \leftrightarrow \tau_2$.}
\be \label{ccpartial}
\p_{\R,\,{[c,\,c]}}^{\rm 1-loop}(p) \cong\, -8\Mp^ 2  |\Delta\eta|^2 \p_\R^{\rm tree}(p) \int^{\taue}_{\taus} d\tau_1 \int^{\tau_1}_{\tau_s}d\tau_2\, a^2(\tau_2)\epsilon(\tau_2) \int \frac{d\bk}{(2\pi)^3}\, \mathrm{Im}\left(\mathcal{X}_k\mathcal{Y}_{\q}  \right),
\ee
where
\be
\mathcal{X}_k= \R'_k(\tau_2)\R'^*_k(\tau_1),\qquad \mathrm{and}\qquad  \mathcal{Y}_{k} = \R'_{k}(\tau_2)\R^*_{k}(\tau_1).
\ee
As before, we replace commutators such as $[\Rh_{1}',\Rh_{\bp}]$ with the Wronskian conditions in Eq. \eqref{approxcomm2}. 
From Eq. \eqref{ccpartial} we single out terms which depend only on $\tau_2$ and we carry out integration by parts using the linear equation of motion for $\R$, i.e. $(a^2\epsilon\R')'= a^2\epsilon\partial^2\R$. As mentioned, we do not need to use the explicit form of the mode functions.
Proceeding in this way, the integrals over $\tau_2$ can be rewritten as
\begin{align}\label{parts}
\int^{\tau_1} d\tau_2 a(\tau_2)^2 \epsilon(\tau_2) \R'_k\R'_{\q} 
= (a^2 \epsilon  \,\R'_k\R_{\q})|_{\tau_1} + \int^{\tau_1} d\tau_2 a^2 \epsilon  \,k^2\R_k\,\R_{\q}
\end{align}
We symmetrize the integrand over the two momenta $k$ and $|\bp-\bk|$, then we insert the previous expression and its complex conjugate in \eqref{ccpartial}. Surface terms cancel and after regrouping what is left, we finally obtain
\begin{align}\nonumber
\p_{\R,\,{[c,\,c]}}^{\rm 1-loop}(p)\cong  \,& -4 \Mp^ 2 |\Delta\eta|^2 \p_\R^{\rm tree}(p) \int^{\taue}_{\taus} d\tau_1 \int^{\tau_1}_{\tau_s}d\tau_2 a^2(\tau_2)\epsilon(\tau_2) \int \frac{d\bk}{(2\pi)^3}\,  (k^2+|\bp-\bk|^2)\\ \label{finalcc}
&\qquad\,\, \times\,\, \mathrm{Im}\left( \R^*_k(\tau_1) \R'^*_{\q}(\tau_1)\, \R_k(\tau_2) \R_{\q}(\tau_2)   \right).
\end{align}

\subsubsection{[d-c] contribution}
Let us now consider the contribution in Eq. \eqref{dc}. By inserting the expression for the two Hamiltonian operators, one gets:
\begin{align}\nonumber
\langle \Rh_{\bp}\Rh_{\bp'}\rangle_{[d,\,c]} &= -\Mp^ 4\frac{|\Delta\eta|^2}{2}\int^{\taue}_{\taus} d\tau_1 (a^2\epsilon)|_{\tau_1} \int^{\tau_1}_{\tau_s} d\tau_2(a^2\epsilon)|_{\tau_2} \\ \label{dc2}
&\quad \times \int d\bK \cdot \mathcal{A}\, \langle [(\Rh_{\bk_1}\Rh_{\bk_2}\Rh_{\bk_3})|_{\tau_2},[(\Rh'_{\bk_4}\Rh'_{\bk_5}\Rh_{\bk_6})|_{\tau_1},\Rh_{\bp}\Rh_{\bp'}]] \rangle.
\end{align}
where
\be\label{pref}
\mathcal{A} = - \frac{1}{3}\sum^3_{i=1}k^2_{2,i}.
\ee
The first nested commutator is the same as in the previous section, we thus use Eq. \eqref{firstun} and neglect tadpoles. Using the same compact notation as defined in Eq. \eqref{compact} and the same approximation of the previous sections we have 
\begin{align}\nonumber
[\Rh_2 \Rh_2 \Rh_2,[\Rh_1'\Rh_1'\Rh_1,\Rh_{\bp}\Rh_{\bp'}]]&=4[\Rh_1',\Rh_{\bp'}]\cdot [\Rh_2 \Rh_2 \Rh_2,\Rh_1'\Rh_1]\Rh_{\bp}\\ \nonumber
&= 4[\Rh_1',\Rh_{\bp'}]\cdot [\Rh_2,\Rh_1](2 \Rh_2\Rh_2\Rh_1' \Rh_{\bp} + \Rh_2 \Rh_{\bp}[\Rh_1',\Rh_2] + \Rh_1'\Rh_2\Rh_2\Rh_{\bp})\\ \label{inter}
\,&\,\,+ 4[\Rh_1',\Rh_{\bp'}]\cdot [\Rh_2,\Rh_1'](2\Rh_2\Rh_2\Rh_1\Rh_{\bp}+ \Rh_2\Rh_{\bp}[\Rh_1,\Rh_2]+\Rh_1\Rh_2\Rh_2\Rh_{\bp}).
\end{align}
As before, once replaced by Eq. \eqref{approxcomm2}, the commutator $[\Rh_1',\Rh_{\bp'}]$ annihilates the prefactor with the $\tau_1$ dependence.
Inserting \eqref{inter} in Eq. \eqref{dc2}, and going back to the dimensionless power spectrum, we obtain
\begin{align}
\p_{\R,\,{[d,\,c]}}^{\rm 1-loop}(p) \cong -\Mp^ 2\frac{|\Delta\eta|^2}{2}\p_\R^{\rm tree}(p)\int^{\taue}_{\taus} d\tau_1 \int^{\tau_1}_{\tau_s}(a^2\epsilon)|_{\tau_2} d \tau_2 \int \frac{d \bk}{(2\pi)^3} \cdot \mathcal{A} \cdot 3\cdot 8 \,\mathrm{Im}\left(\mathcal{Z}_k  \mathcal{W}_{\bq}  \right) 
\end{align}
where
\be
\mathcal{Z}_k = \R_k(\tau_2)\R_k'^*(\tau_1),\quad \mathrm{and}\quad \mathcal{W}_k = \R_{k}(\tau_2)\R^*_{k}(\tau_1).
\ee
After integrating over the various momenta the prefactor in Eq. \eqref{pref} becomes\footnote{We would have found terms proportional to $p^2$ also in the previous section by including terms proportional to $\R_p'$ (which we always neglect) and doing integration by parts as in Eq. \eqref{parts}.}
\be
\mathcal{A} = - \frac{1}{3}(k^2 + |\bp-\bk|^2 + p^2)\simeq - \frac{1}{3}(k^2 + |\bp-\bk|^2),
\ee
it is then easy to rewrite the final expression for this contribution as
\begin{align}\nonumber
\p_{\R,\,{[d,\,c]}}^{\rm 1-loop}(p) \cong\,\,  \,& 4 \Mp^ 2 |\Delta\eta|^2 \p_\R^{\rm tree}(p)\int^{\taue}_{\taus} d\tau_1 \int^{\tau_1}_{\tau_s}d\tau_2 a^2(\tau_2)\epsilon(\tau_2) \int \frac{d\bk}{(2\pi)^3}\,  (k^2+|\bp-\bk|^2)\\ \label{finaldc}
&\qquad\,\, \times\,\, \mathrm{Im}\left( \R^*_k(\tau_1) \R'^*_{\q}(\tau_1)\, \R_k(\tau_2) \R_{\q}(\tau_2)   \right).
\end{align}
By summing Eq. \eqref{finalcc} with Eq. \eqref{finaldc}, one finds the desired results:
\be
\p_{\R,\,{[c,\,c]}}^{\rm 1-loop}(p)+\p_{\R,\,{[d,\,c]}}^{\rm 1-loop}(p) \cong 0,
\ee
where, as before, the symbol $\cong$ identifies quantities up to volume suppressed terms and up to term proportional to the Green's function in Eq. \eqref{approxcomm}.
\section{Conclusions}
One-loop corrections on large scales from short modes which are enhanced due to a period of non-slow-roll evolution may lead to contributions independent on the ratio between the two scales involved in the problem. That is quite counter-intuitive. In the present study, we first hint about the absence of these type of corrections once including all relevant diagrams. 

We have shown the relevance of apparently irrelevant total derivative terms. These provide non-volume suppressed contributions of equal type (and opposite sign) as the one founded by including only bulk operators. In particular, we explicitly demonstrate that the leading contribution in Eq. \eqref{first}, identified in previous analyses, precisely vanishes when a total derivative term from the cubic action for $\zeta$ is taken into account. We use two equivalent forms of the cubic interaction Hamiltonian for the comoving curvature perturbation---see Eqs. \eqref{Option1}-\eqref{Method2}. Each method yields valuable lessons. Our first approach shows that, in general, boundary terms cannot be neglected in this context. Our second method confirms the explicit cancellation of the non-volume suppressed terms in Eq. \eqref{first} without the need of doing any explicit time integrals. This indicates that our result is quite general, i.e. it goes far beyond the specific setup of a sharp transition during an ultra-slow-roll phase, which was the original motivation behind this computation.

There are several possible extensions that one may envisage for future studies. Let us just list a few. The first step would be to evaluate how all non-volume suppressed terms add up when including the relevant part of the quartic Hamiltonian (see \cite{Firouzjahi:2023aum} for an initial attempt at this), and without neglecting contributions proportional to the Green's function in Eq. \eqref{approxcomm}. This is the content of our follow-up analysis in \cite{Fumagalli:2024jzz}; see also \cite{Kawaguchi:2024rsv}. Further, by not taking the long-scale short-scale approximation, one can generalize the computations above and derive one-loop effects on all wavelengths as induced by enhanced short modes. A physical effect will likely appear around momenta corresponding to the short scales and a proper renormalization procedure should be implemented there. Further, we focus our attention on 1PI diagrams and neglect tadpoles. The latter should also be computed and properly regularized in this context. Moreover, as already mentioned, it would be interesting to prove the cancellation of the would be leading terms by implementing the field redefinition procedure in \cite{Maldacena:2002vr} adapted to this framework. We leave all this for future investigations.

\section*{Acknowledgments}
The author sincerely thanks Jaume Garriga for a first discussion on this issue and for numerous subsequent enlightening conversations. 
The author is also greatful to Cristiano Germani, Sadra Jazayeri, Lucas Pinol and Alexandre Serantes for different and useful inputs and clarifications.
The research of J.F. is supported by the State Agency for Research of the Spanish Ministry
of Science and Innovation through the “Unit of Excellence María de Maeztu 2020-2023” award to the Institute of Cosmos Sciences (CEX2019-000918-M) and by grants 2021-SGR00872 and PID2019-105614GB-C22.


\bibliographystyle{JHEP}
\bibliography{Biblio}
\end{document}